\documentclass{aa}

\usepackage{amsmath,txfonts,graphicx,natbib,epsfig,amssymb,lscape}
\bibpunct{(}{)}{;}{a}{}{,}

\def\deg{\hbox{$^\circ$}}
\def\micron{\hbox{$\mu$m}}
\def\Teff{\hbox{$T_{\rm eff}$}}

\begin{document}

\title{Stellar populations in the Galactic bulge \thanks{Figure 9 is
    only available in electronic form via http://www.edpsciences.org}}
\subtitle{Modelling the Galactic bulge with TRILEGAL}
\author{E. Vanhollebeke \inst{1} \and 
M.A.T. Groenewegen \inst{2,1} \and
L. Girardi \inst{3}} 

\institute{
Instituut voor Sterrenkunde, K.U.Leuven, Celestijnenlaan 200D, B--3001 Leuven, Belgium 
\and Royal Observatory of Belgium, Ringlaan 3, B--1180 Brussels, Belgium
\and  Padova Osservatorio Astronomia, Vicolo dell'Osservatorio 5, I--35122 Padova, Italy}

\date{Received / Accepted }
\abstract
{}
{The aim of this paper is to study the characteristics of the stellar
  populations and the metallicity distribution in the Galactic
  bulge. We study the entire stellar population, but also retrieve
  information using only the red clump stars.}
{To study the characteristics of the stellar populations and the
  metallicity distribution in the Galactic bulge, we compared the
  output of the galaxy model TRILEGAL, which implements the Binney et
  al. (1997) bulge 
  model, with observations from 2MASS and
  OGLE-II. A minimisation procedure has been set up to retrieve the
  best fitting model with different stellar populations and
  metallicity distributions.}
{Using the TRILEGAL code we find that the best model resembling the
  characteristics of the Galactic bulge is a model with the distance
  to the Galactic centre $R_0 = 8.7\pm^{0.57}_{0.43}$~kpc, the major
  axis ratios of the bar $1:\eta:\zeta = 1 : 0.68\pm_{0.19}^{0.05} :
  0.31\pm_{0.04}^{0.06}$, and the angle between the Sun-centre line
  and the bar $\phi = 15\deg\pm_{12.7}^{13.3}$. Using these
  parameters the best model is found for a burst of 8 Gyr, although it
  is almost indistinguishable from models with ages of 9 and
  10~Gyr. The metallicity distribution found is consistent with
  metallicity distributions in the literature based on spectroscopic results.}
{}

\keywords{Galaxy: bulge -- Galaxy: structure -- Stars: distances --
  Galaxy: fundamental parameters -- Stars: Hertzsprung Russel (HR) and
  C-M diagrams}


\maketitle


\section{Introduction}
\label{section introduction}

The central part of our Milky Way Galaxy (MWG) consists of the
Galactic bulge (GB), the central part of the halo, and the Galactic
centre (GC). Over the years, several studies have tried to understand
the characteristics of these parts of the MWG, for example about the
formation history, the rate of star formation, the relations between
different stellar components, and its kinematics \citep{Wyse1995}.\\

Understanding the formation history of galactic spheroids
(i.e. elliptical galaxies and the bulges of spiral galaxies) is
crucial in understanding galaxy formation in general as these
spheroids contain a large fraction of all stellar mass in the local
universe \citep[and references therein, hereafter ZRO2003]{Zoccali2003}. 
Because of its proximity, our GB is the spheroid for which the most
information is accessible. Issues that can be addressed in the GB are,
e.g., the initial mass function, the distribution of stellar ages, and
the 3D structure. For more distant spheroids, this sort of information
is only accessible by very indirect means based on the study of the
integrated light.\\

Many different approaches exist to studying galaxy formation. Often
galaxy models are used to study the entire content of a galactic
component such as the bulge \citep{Dwek1995, Binney1997,
  Freudenreich1998, Bissantz2002, Robin2003, Merrifield2004}, but
studies of the same kinds of stars can also reveal a wealth of
information. Examples of this are, e.g., the use of red clump stars
\citep{Stanek1997, Paczynski1998, Udalski1998, Babusiaux2005,
  Nishiyama2006, Lopez-Corredoira2007, Rattenbury2007}, RR~Lyrae stars
\citep{Fernley1987, Feast1997a, Udalski1998}, masers \citep{Reid1988,
  Sevenster1999a}, and Mira variables \citep{Whitelock1992,
  Groenewegen2005}. Our target is both to perform a calibration of the
bulge geometry and to explore the distribution of ages that can be
inferred from these data.\\

The paper is organised as follows. In Sect.~\ref{section GB} we give a
resume on the models and parameters generally used to study the GB,
the star formation rate, and the metallicity distribution of the GB. In
Sect.~\ref{section trilegal} we give a brief introduction to the
galaxy model used in this work. Section~\ref{section observational data}
is dedicated to the observational data used. Section~\ref{section
  analysis} explains the method used and Sect.~\ref{section results}
discusses the results derived using this method. Section~\ref{section RC
  stars} discusses the results using the same method if only the
  red clump stars are used. In Sect.~\ref{section conclusions} we
summarise and come to the conclusions.\\


\section{The Galactic bulge}
\label{section GB}

At the moment, it is still not clear whether one should draw a
distinction between a triaxial central bulge and a bar component
\citep{Merrifield2004}. Sometimes the bulge is referred to as the bar
\citep[see e.g.][]{Sevenster1999a} and in other cases studies find a
bar within the bulge \citep[see e.g.][]{Lopez-Corredoira2007,
  Cabrera-Lavers2007}. In this paper we will adopt the formalism
described by \citet{Sevenster1999b}: we will use the term
\textit{Galactic bulge} for the Galactic component seen in the general
direction of $|\ell| \le 10\deg$, we will use \textit{bar} if we refer to
the prolate or triaxial component of the GB.\\

This section gives a brief overview of recent studies proposing
different models and parameters describing the GB. Also the possible
star formation rate and age metallicity relations will be described.\\

\subsection{Models describing the Galactic bulge and its bar}
\label{section binney model}

Many different geometries can be assumed for the GB. Oblate
spheroidal models can be a good start, providing a first order
estimate of the bulge luminosity and minor to major axis ratios
\citep{Dwek1995}. As became clear over the past decades, there is a
triaxial part, the bar, which needs a triaxial model. There are three
types of triaxial models proposed: Gaussian-type functions,
exponential-type functions, and power-law functions \citep[see][and
  references therein]{Dwek1995}. \citet{Dwek1995} find that the best
model to fit the COBE/DIRBE data for the bulge is an exponential-type
function with a modified spheroid included. \citet{Robin2003} use the
same parameter set as \citet{Dwek1995} introduced to fit the
bulge. They used a boxy Gaussian type function, but did not model the
inner parts of the bulge. \citet{Binney1997} and also
\citet{Bissantz2002} used a truncated power-law to model the bulge. In
this paper, we will use the \citet{Binney1997} formalism to model the
number density of stars $f_{\rm bulge}$, which is given by:
\begin{eqnarray}
f_{\rm bulge} & = & f_0 \frac{e^{-a^2/a_m^2}}{(1+a/a_0)^{1.8}} {\ \rm and}\\
a & = & \left( x^2 + \frac{y^2}{\eta^2} +
\frac{z^2}{\zeta^2}\right)^{1/2}. 
\end{eqnarray}
In these formulae, $a_m$ represents the scale length, $a_0$ the inner
truncation length, and $f_0$ is the stellar density per
kpc$^{3}$ of the bulge. The axis-ratios of the bar are
given by: $1:\eta:\zeta$.\\

\subsection{Parameters describing the Galactic bulge}
\label{section parameters}
  \begin{table*}
    \caption{A not complete overview of recent values of parameters
      describing the GB, its bar and the distance to the
      GC. \label{table parameters}}
    \centering
    \tabcolsep=3pt
    \begin{tabular}{lllllll}
      \hline\hline
      Reference & $R_0$ & $a_m$ & $a_0$ & 1 : $\eta$ : $\zeta$ & $\phi$ &
      Based on \\
      & [kpc] & [kpc] & [pc] & & [\deg] & \\
      \hline
      \citet{Fernley1987} & 8.0 $\pm$ 0.65 & & & & & RR~Lyrae stars\\
      \citet{Reid1988} & 7.1 $\pm$ 1.5 & & & & & H$_2$O maser spots\\
      \citet{Whitelock1992} & 9.1 & & & 1:0.25:0.25 & 45 & Mira   variables \\
      \citet{Dwek1995} & & & & 1:0.33$\pm$0.11:0.23$\pm$0.08 & 20 $\pm$
      10 & \textit{COBE}/DIRBE\\ & & & & & & surface brightness map\\
      \citet{Binney1997} & & 1.9 & 100 & 1:0.6:0.4 & 20 &   \textit{COBE}/DIRBE\\
      & & & & & & surface brightness map\\
      \citet{Feast1997a} & 8.1 $\pm$ 0.4 & & & & & RR~Lyrae stars\\ 
      \citet{Stanek1997} & & & & 1:0.43:0.29 & 20 -- 30 & red clump stars\\
      \citet{Freudenreich1998} & & 2.6 & & & & DIRBE full-sky surface\\
      & & & & & & brightness map\\
      \citet{Paczynski1998} & 8.4 $\pm$ 0.4 & & & & & red clump  stars\\
      \citet{Udalski1998} & 8.1 $\pm$ 0.15 & & & & & RR~Lyrae stars\\
      \citet{Udalski1998} & 8.1 $\pm$ 0.06 & & & & & red clump stars\\
      \citet{Sevenster1999a} &  & 2.5 & & & 44 & OH/IR stars \\  
      \citet{Bissantz2002}  & & 2.8 & 100 & 1:(0.3 -- 0.4):0.3 & 20 -- 25 & 
      \textit{COBE}/DIRBE\\ & & & & & & $L$-band map\\
      \citet{Eisenhauer2003} & 7.94 $\pm$ 0.42 & & & & & stars  orbiting black hole\\
      \citet{Robin2003} & & & & 1:0.27:0.27 & 11.1 $\pm$ 0.7 &  Hipparcos data \\
      \citet{Merrifield2004} & & & & 1:0.6:0.4 & 25 & \ion{H}{i} gas  and \textit{COBE}/DIRBE\\
      & & & & & & surface brightness map \\
      \citet{Babusiaux2005} & 7.7 $\pm$ 0.15 & & & & 22 $\pm$ 5.5 &  red clump stars\\
      \citet{Eisenhauer2005} & 7.62 $\pm$ 0.32 & & & & & stars  orbiting black hole\\
      \citet{Groenewegen2005} & 8.8 $\pm$ 0.4 & & & & 47 & Mira variables\\
      \citet{Lopez-Corredoira2005} & & & & 1:0.5:0.4 & 20 -- 35 & 2MASS star counts \\
      \citet{Nishiyama2006} & 7.51 $\pm$ 0.10 $\pm$ 0.35 & & & & & red  clump stars\\
      \citet{Lopez-Corredoira2007} & & & & & 43 & red clump stars \\
      \citet{Rattenbury2007} & & & & 1:0.35:0.26 & 24 -- 27 & red clump stars\\
      \citet{Ghez2008} & 8.0 $\pm$ 0.4 & & & & & stars orbiting black hole\\
      \citet{Gillessen2009}  & 8.33 $\pm$ 0.35 & & & & & stars orbiting black hole \\
      \hline
    \end{tabular}
  \end{table*}

The typical parameters needed to describe the GB (see previous
section) are the scale length ($a_m$) and inner truncation length
($a_0$) of the bulge, the axis-ratios (1~:~$\eta$~:~$\zeta$) of the
bar, the angle between the Sun-centre line and the major axis of the
bar ($\phi$), and the scaling parameter ($f_0$). Of course also the
Sun's distance to the Galactic centre ($R_0$) is an important
parameter in this context.\\

One of the most studied types of stars in the GB are the red clump
stars. The red clump stars are considered as standard candles to
determine distances. They can also be used to estimate the axis ratios
of the bulge and the angle $\phi$. Using these red clump stars, values
between 7.4 and 8.4~kpc for $R_0$ have been found \citep[see
  e.g.][]{Paczynski1998, Udalski1998, Babusiaux2005,
  Nishiyama2006}. Values for $\phi$ derived from red clump stars are
situated either around 20\deg\ or around 45\deg\ (see
e.g. \citet[$\phi=20\deg - 30\deg$]{Stanek1997}, \citet[$\phi = 22\deg
  \pm 5.5\deg$]{Babusiaux2005}, and
\citet[$\phi=43\deg$]{Lopez-Corredoira2007}). Recent work on the red
clump stars in the GB comes from \citet{Cabrera-Lavers2007} and
\citet{Rattenbury2007}. \citet{Cabrera-Lavers2007} claim two very
different large-scale triaxial structures in the inner Galaxy. A first
component is a long thin stellar bar ($|b|<2\deg$) with a position
angel of $43.0\deg \pm 1.8\deg$. The second component is a distinct
triaxial bulge that extends to at least $|b|\le 7.5\deg$ and has a
position angle of 12.6\deg $\pm$ 3.2\deg. \citet{Rattenbury2007} study
the red clump stars observed with OGLE in the OGLE-II phase and find
that the bar's major axis is oriented at 24\deg~--~27\deg\ to the
Sun-Galactic centre line-of-sight.\\ Another kind of stars that
is considered a standard candle are the RR~Lyrae stars. Using these
stars typical values for $R_0$ of $\sim$~8.0~kpc have been found
\citep[see e.g.][]{Fernley1987, Feast1997a, Udalski1998}.\\ Other
types of stars that have been studied to characterise the GB are
e.g. Mira variables (see e.g. \citet[$R_0 = 8.8$~kpc and $\phi =
  43\deg$]{Groenewegen2005} and \citet[$1:\eta:\zeta = 1:0.25:0.25$,
  $\phi=45\deg$]{Whitelock1992}), H$_2$O masers \citep[see][$R_0 =
  7.1$~kpc]{Reid1988}, and OH/IR stars \citep[see][$a_m =
  2500$~pc and $\phi = 44$\deg]{Sevenster1999a}.\\

Studying one type of star is not the only method to obtain information
on the characteristics of the Galactic bulge and distance to the
Galactic centre. \citet{Eisenhauer2003} and \citet{Eisenhauer2005}
studied stars orbiting the central black hole and found values for
$R_0$ of respectively $7.94 \pm 0.42$~kpc and $7.62 \pm
0.32$~kpc. Recently, \citet{Ghez2008} and \citet{Gillessen2009} 
used additional data to find, respectively, $8.0 \pm 0.4$ kpc and
  $8.33 \pm 0.35$ kpc.  Also surveys like 2MASS and the surface
brightness maps of \textit{COBE}/DIRBE were used. Studies based on
these databases find values for $\phi$ around 20\deg\ 
\citep[see e.g.][]{Dwek1995, Binney1997, Bissantz2002,
  Lopez-Corredoira2005}, scale lengths $a_m$ between 1900~pc and
2800~pc \cite[see e.g.][]{Binney1997,Freudenreich1998, Bissantz2002},
and an axis-ratio of 1:0.5-0.6:0.4
\citep{Merrifield2004,Lopez-Corredoira2005}, except \citet{Dwek1995}
who find lower values for the axis-ratio
(1:0.33:0.23). \citet{Robin2003} used HIPPARCOS data to model and
retrieved axis-ratios of 1:0.27:0.27 and an angle $\phi$ of
11.1\deg~$\pm$~0.7\deg.\\

The Milky Way Galaxy consists not only of stars, there is also an
amount of gas present. The location of \ion{H}{i} gas in the galaxy
\citep[see Fig.~2 in][]{Merrifield2004} immediately reveals the
non-axisymmetric distribution of the gas. This non-axisymmetric
distribution can also be seen when one measures the velocities of the
interstellar CO molecule \citep{Dame2001}. Using the distribution of
the \ion{H}{i} gas \citet{Merrifield2004} retrieve an angle $\phi$ of 25\deg.\\

Table~\ref{table parameters} gives an overview of these studies and
the parameters retrieved. Concerning the angle between the Sun-centre
line and the bar $\phi$, Table~\ref{table parameters} clearly shows two
groups of values: a low value around $20\deg$ and a higher value
around $45\deg$. According to \citet{Sevenster1999a} the lower values
found for $\phi$ arise when the longitude range used is too narrow or
when low latitudes are excluded. \citet{Groenewegen2005} point out
that this could also be due to the fact that these studies trace
different populations, which may be distributed differently.\\

\subsection{Star formation rate and metallicity distribution}
\label{section SFR and AMR}

Besides the uncertainty in the literature on the parameters describing the
GB geometry and its bar, the characteristics of the stars located
in the GB are being debated.\\

Lately, studies show traces of an intermediate age population
in the GB. Before, it was believed that the GB consisted only of an
old population (more than 10~Gyr). Recent studies still reveal that
the GB is dominated by this old population, but traces of a smaller
intermediate age population can no longer be ignored.\\
\citet{Holtzman1993} studied a field in Baade's Window and concluded
that, based on the luminosity function, there exist not only old stars
in the GB, but also a significant, although unquantified, number of
intermediate age stars (less than 10~Gyr). \citet{Ortolani1995}
conclude that there is no age difference between the majority of bulge
stars and the halo globular clusters and no more than $\sim 10\%$ of
the bulge population can be represented by intermediate age stars
\citep[see also][ZRO2003]{Ortolani2001}. \citet{Feltzing2000}
found no significant young stellar population in the GB, but emphasise
that it is still possible to have an age range of several
Gyr. \citet[hereafter vLGO2003]{vanLoon2003} found that in addition to
the dominant old
population ($\ge 7$~Gyr), there is also an intermediate-age population
($\sim 200$~Myr~--~7~Gyr) and possibly even younger ($\le 200$~Myr)
stars are found across the inner bulge (although the latter could be
attributed to foreground stars). Concerning an intermediate age
population, \citet{Groenewegen2005} found that the Mira variables in
the studied OGLE bulge fields have ages of a few Gyr. Also
\citet{Uttenthaler2007} found some AGB stars that might be tracers of
a younger population: some of their selected AGB stars in the GB show
Technetium, which indicates that these AGB stars must originate from a
younger population.\\

The detection of a metallicity spread or absence of it and the
metallicity distribution reveals information on the formation
history. During the last decade, there were several papers studying
the metallicity in the GB. \citet{McWilliam1994} obtained for 14 M
giants in the GB high resolution spectroscopy. They found a mean
metallicity of $<$[Fe/H]$> = -0.19 \pm 0.02$.  \citet{Sadler1996}
studied K and M giants in Baade's Window and found a mean abundance of
$<$[Fe/H]$> = -0.11 \pm 0.04$. \citet{Feltzing2000} studied HST images
and found that the metallicity of the bulge is equal to that of the
old disk and that there is only a marginal evidence for a central
metallicity gradient. Also \citet{Ramirez2000} studied M giants in the
GB. They found a mean metallicity $<$[Fe/H]$> = -0.21 \pm 0.30$ and no
evidence for a metallicity gradient along the minor or major axes of
the inner bulge ($R < 560$~pc). vLGO2003 found a mean metallicity
[M/H]~$\sim 0.5$, but also stars with a metallicity of +0.5 and -2 are
common, but not dominant. The old stars in the population ($\ge
7$~Gyr) tend to have higher metallicities (see their Fig.~18). They
propose that the metallicity distribution of the old stars might be
bimodal: one component of super-solar metallicity and another of
subsolar metallicity. ZRO2003 found that most of the bulge stars have
a metallicity [M/H] between +0.1 and -0.6 with a peak at -0.1.\\


\section{TRILEGAL}
\label{section trilegal}

In this paper, we use the code TRILEGAL \citep[TRIdimensional modeL of
thE GALaxy,][]{Girardi2005}, a population synthesis code to
simulate the stellar content towards any direction on the sky. In this
paper the model will be used to compute colour-magnitude diagrams
(CMD) towards the GB. This means that we simulate the
photometric properties of stars located towards a given direction
$(\ell,b)$ and complete down to a given limiting magnitude.\\

Fig.~1 in \citet{Girardi2005} shows the general scheme of the 
TRILEGAL code. As is shown in their Fig.~1, the input to the model
consists of four main elements:
\begin{itemize}
\item a library of stellar evolutionary tracks. It is essentially
  composed of the \citet{Girardi2000} tracks with some updates at low
  metallicities (Girardi 2002, unpublished), and now complemented with
  the TP-AGB models by \citet{Marigo2007}. These latter are expected
  to describe the TP-AGB phase with much more realistic lifetimes and
  luminosities than in the original \citet{Girardi2000} isochrones.\\
\item a library of synthetic spectra, which is used to compute bolometric
corrections and extinction coefficients for the simulated stars. They
are described in detail by \citet{Girardi2002}.\\
\item the instrumental setup, describing the settings of the telescope
  to be used, e.g. filters, detectors, and the effective sky area to be
  simulated.\\
\item a detailed description of the Galaxy components, being the
  Galactic thin and thick disk, the Halo and the Bulge. For each
  component a Star Formation Rate (SFR), Age--Metallicity Relation
  (AMR), and space densities are assumed. These parameters are
  different for every Galaxy component, therefore the Galactic
  components are treated separately by the code. Also the Initial Mass
  Function (IMF) and interstellar absorption belong to this group of
  input parameters.\\
\end{itemize}

The TRILEGAL code itself performs a Monte Carlo simulation to generate
stars in accordance with the input (see above). The SFR,
AMR and IMF define a stars age, metallicity, and mass. An
interpolation in the grids of evolutionary tracks
results in the absolute photometry, which
is then converted to apparent magnitudes based on the bolometric
corrections, distance modulus, and extinction.\\

The output of the TRILEGAL code is a catalogue of stars that contains
for each star the Galactic component in which they are located, stellar
age, metallicity [M/H], initial mass, luminosity $L$, effective
temperature \Teff, gravity $g$, distance modulus $m-M_0$, visual
extinction $A_V$, core mass and the \emph{perfect} photometric data in
each of the wanted filters. To be able to compare the output in
CMDs with observational data one has to perform some extra
calculations which are not included in the TRILEGAL code such as
adding photometric errors, saturation of the bright stars in real
observations, and completeness.\\

A more detailed description of the TRILEGAL code and the parameters
used concerning the disk and halo can be found in \citet{Girardi2005}.\\

Extinction is calculated by the TRILEGAL code for each object
  separately based on the input $A_V$ value and the distance modulus of
  the objects. Therefore, the extinction varies along the line-of-sight
  \citep[see][for a more detailed description on the way the
  extinction is distributed]{Girardi2005}. Some changes are
  made, however, to this part of the code with respect to
  \citet{Girardi2005}. Not only can we now give the extinction value
$A_V$ as an input parameter to the code, but a spread on the $A_V$ has
also been used in this work. The extinction values and their sigma
used can be found in Table~\ref{table selected fields} and are based
on \citet{Sumi2004}.\\

An important point is that we have used TRILEGAL with the original
filter curves provided for 2MASS \citep{Skrutskie2006} and OGLE-II
\citep{Udalski1997}, so simulating as far as possible the original
photometric systems\footnote{Our experience is that the simulation of
the 2MASS photometric system presents zero-point offsets of just a few
0.01~mag. Zero-point errors of similar magnitude are just expected in
synthetic photometry applied to present-day libraries of synthetic
spectra, and are also likely present in our simulations of the OGLE-II
system.}. The bolometric corrections for 2MASS have already been
described in detail by \citet{Bonatto2004}. Those for OGLE-II have
been newly computed, and apart from being used in TRILEGAL have also
been incorporated into the Padova database of isochrones. Extensive
tables of bolometric corrections and isochrones in the OGLE-II system
are now made available at the websites {\tt
http://pleiadi.pd.astro.it/isoc\_photsys.02} (static tables) and {\tt
http://stev.oapd.inaf.it/cmd} (interactive web form).\\

Furthermore, extinction coefficients in the several 2MASS and OGLE-II
passbands have been computed for a G2V star using the
\citet{Cardelli1989} extinction curve with $R_V=3.1$. The extinction
coefficients turn out to be $A_{U_{\rm OGLE}}/A_V=1.524$, $A_{B_{\rm
    OGLE}}/A_V=1.308$, $A_{V_{\rm OGLE}}/A_V=1.021$, $A_{I_{\rm
    OGLE}}/A_V=0.571$, $A_{J_{\rm 2MASS}}/A_V=0.290$, $A_{H_{\rm
    2MASS}}/A_V=0.183$, and $A_{K_{\rm s,2MASS}}/A_V=0.118$.\\

These numbers would be suitable to simulate extinction in the nearby disk,
but do not apply to a bulge with an anomalous ratio of total to
selective extinction \citep[see][]{Udalski2003}. Since the reasons for such
anomalous ratios are still not understood, one does not know really
how to properly simulate extinction for the bulge. Our choice has been
very pragmatic: we assume that the generalised \citet{Cardelli1989}
extinction curve still applies, and just change its input $R_V$ value
until the mean $R_{VI}=1.964$ value for the bulge \citep{Sumi2004} is
recovered. Therefore, we find that the anomalous extinction is
reproduced with a $R_V=2.4$, which produces $A_{U_{\rm
OGLE}}/A_V=1.715$, $A_{B_{\rm OGLE}}/A_V=1.386$, $A_{V_{\rm
OGLE}}/A_V=1.017$, $A_{I_{\rm OGLE}}/A_V=0.506$, $A_{J_{\rm
2MASS}}/A_V=0.256$, $A_{H_{\rm 2MASS}}/A_V=0.162$, $A_{K_{\rm
s,2MASS}}/A_V=0.101$. These values are assumed in this work.\\


\section{Observational Data}
\label{section observational data}

\subsection{Two Micron All Sky Survey -- 2MASS}
\label{section 2mass}
The Two Micron All Sky Survey (2MASS) project is a collaboration
between The University of Massachusetts and the Infrared Processing
and Analysis Center (JPL/Caltech). The resulting 2MASS project made
uniformly-calibrated observations of the entire sky (with a sky
coverage of 99.998\%) in the $J$ (1.24~\micron), $H$ (1.66~\micron)
and $K_s$ (2.16~\micron) near-infrared bands \citep{Skrutskie2006}.\\ 

In this paper, we will use the 2MASS Point Source Catalog (PSC), which
contains astrometry and photometry in the three survey bandpasses for
470~992~970 sources. For every source, position, magnitude,
astrometric and photometric uncertainties, and flags, indicating
the quality of the source characterisations, are provided.\\

\subsection{Optical Gravitation Lensing Experiment -- OGLE}
\label{section ogle}
The Optical Gravitational Lensing Experiment (OGLE) aims at detecting
dark matter using micro lensing phenomena.  Crowded regions are a
suited place to conduct this survey. The Large Magellanic Cloud (LMC)
was the first target in the second phase. Later on additional targets
were added among 11~deg$^2$ in the GB.\\

In this paper, we will use data from \citet{Udalski2002} from OGLE-II
who present the $VI$ photometric maps of the GB as a natural
by-product of large micro lensing surveys. These maps contain the mean
$VI$-photometry and astrometry of about 30 million stars, covering
about 11~deg$^2$ in different parts of the GB. Each field covers
$14.2' \times 57'$.\\

\subsection{Selected fields}
\label{section selected fields}

\begin{table}
\begin{minipage}[t]{\columnwidth}
  \caption{Properties of the selected fields.}
  \label{table selected fields}
  \centering
  \renewcommand{\footnoterule}{}
  \begin{tabular}{ccccccc}\hline\hline
      Ogle field & $\ell$    & $b$   & $A_V$ \footnote{Based on
      \citet{Sumi2004}} & $\sigma_{A_V}$ $^a$ &
      \multicolumn{2}{c}{\# stars}  \\ 
      (bul\_sc)  &        &       & &     & OGLE          & 2MASS\\
      \hline
      2          &   2.23 & -3.46 & 1.51  & 0.024 & 803~269  & 26~147\\
      7          &  -0.14 & -5.91 & 1.34  & 0.025 & 462~748  & 19~304\\
      8          &  10.48 & -3.78 & 2.14  & 0.042 & 401~813  & 19~411\\
      13         &   7.91 & -3.58 & 2.05  & 0.042 & 569~850  & 20~704\\
      14         &   5.23 &  2.81 & 2.49  & 0.048 & 619~028  & 24~851\\
      17         &   5.28 & -3.45 & 1.92  & 0.031 & 687~019  & 22~241\\
      25         &  -2.32 & -3.56 & 2.33  & 0.042 & 622~326  & 26~328\\
      27         &  -4.92 & -3.65 & 1.69  & 0.025 & 690~785  & 23~623\\
      29         &  -6.64 & -4.62 & 1.53  & 0.024 & 491~941  & 20~692\\
      40         &  -2.99 & -3.14 & 2.92  & 0.066 & 630~774  & 26~169\\
      47         & -11.19 & -2.60 & 2.58  & 0.056 & 300~705  & 20~099\\
      \hline
  \end{tabular}
  \end{minipage}
\end{table}

The fields selected to perform this study are shown in
Table~\ref{table selected fields}. The fields are chosen based on the
OGLE fields from \citet{Udalski2002} and are as much as possible
spread in longitude and latitude around the Galactic centre. The data
from 2MASS has been downloaded according to the positions of the OGLE
fields, using Gator on the IRSA web site
{\tt (http://irsa.ipac.caltech.edu/)}.\\


\section{Analysis}
\label{section analysis}

When calculating a set of models of any kind, one always needs an
objective method to determine why one model is better than
another. One also wants to know the error bars and/or confidence
intervals for the derived parameters.\\

\subsection{Method}
\label{section method}

\begin{figure}[t]
  \begin{center}
    \resizebox{\hsize}{!}{\includegraphics{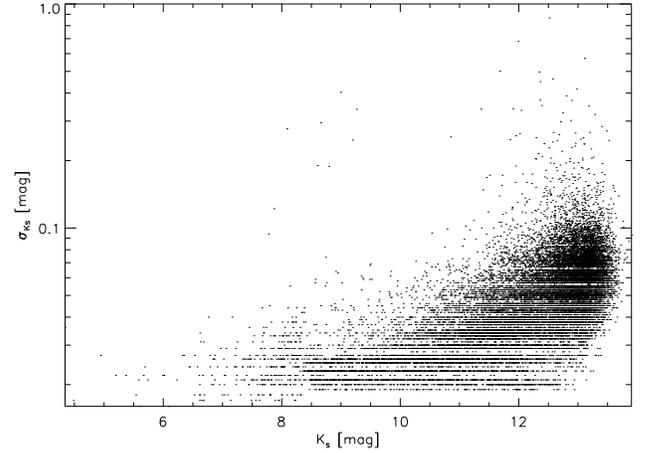}}
    \caption{Example of the photometric errors of the $K_s$-band 2MASS
    data for field bul\_sc40.}
    \label{figure errors 2MASS}
  \end{center}
\end{figure}

The output of the TRILEGAL code consists of \emph{perfect} photometric
data (see Sect.~\ref{section trilegal}). In order to be able to
compare the output of the code with real observations, we need to add
photometric errors to the output of TRILEGAL.\\
Both 2MASS and OGLE-II give individual errors on the
observations. These individual errors are shown in Fig.~\ref{figure errors 2MASS} 
as an example for the 2MASS $K_s$-band data in field bul\_sc40 in
function of the $K_s$-band magnitude. Using this data a density
profile has been generated: for each bin in $K_s$ (size 0.1 mag) we
calculated how likely it is for a star to have an error $\sigma_{K_s}$. 
Each bin in $\sigma_{K_s}$ (size 0.01 mag) is then given a number
between 0 and 1 representing the relative amount of stars in this bin
with respect to all stars with a similar $K_s$-band magnitude. For
each star in the output file of TRILEGAL a random number has been
generated between 0 and 1. The assigned error is then the closest to
$\sigma_{K_s}$ with respect to the assigned density profile.
This method has been used to generate photometric errors in all the
observed magnitude bands from both 2MASS as OGLE-II for each field
individually.\\

\begin{figure}[t]
  \begin{center}
    \resizebox{\hsize}{!}{\includegraphics{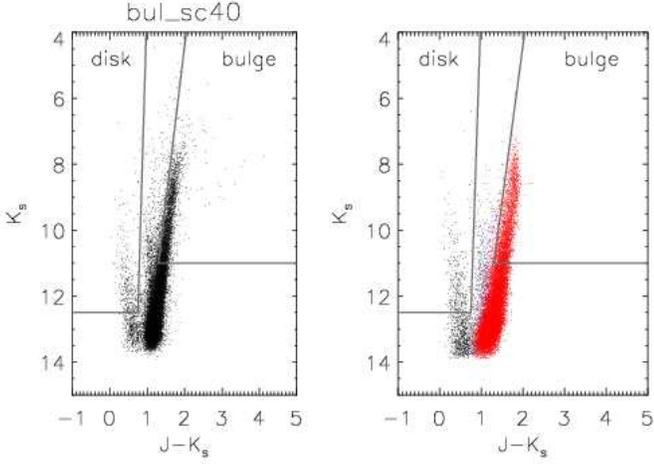}}
    \caption{$J - K_s$ vs $K_s$ CMD for field bul\_sc40. The figure on the
    left are the observations, the figure on the right is a model with
    $R_0 = 8.6$~kpc, $a_m = 2.7$~kpc, $a_0 = 97$~pc, $\eta =
    0.68$, $\zeta = 0.30$, $\phi = 14\deg$, and $f_0 = 427.3$. The
    black dots in the model (right panel) are stars that originate
    from the disk, the dark blue dots originate from the halo and the
    red dots are bulge stars. This colour code will be used in
    all figures of this type. The light grey lines indicate the boxes
    used to compute the model with the observations.}
    \label{figure cmd 2MASS}
  \end{center}
\end{figure}

\begin{figure}[t]
  \begin{center}
    \resizebox{\hsize}{!}{\includegraphics{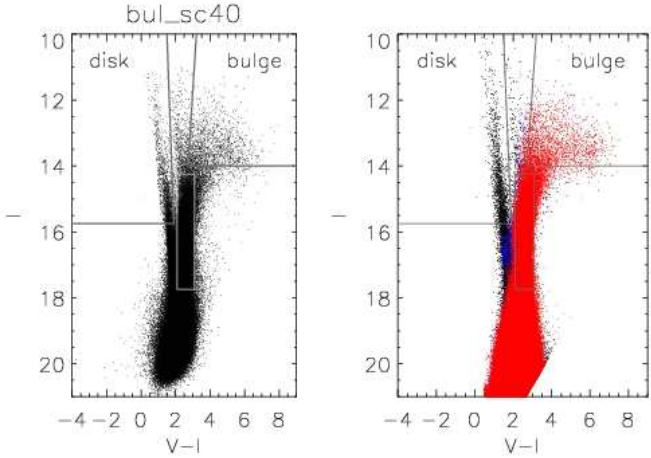}}
    \caption{Same figure and colour code as Fig.~\ref{figure cmd 2MASS} for
    the same model but now for a $V - I$ vs $I$ CMD. The additional box
    which is not seen in Fig.~\ref{figure cmd 2MASS} is used to compare
    the red clump stars.}
    \label{figure cmd OGLE}
  \end{center}
\end{figure}

With the observational data and the data calculated with TRILEGAL,
CMDs have been constructed. For the 2MASS data these CMDs are in $J - K_s$ 
vs $K_s$ (see Fig.~\ref{figure cmd 2MASS}), for the OGLE data the CMDs
are $V - I$ vs $I$ (see Fig.~\ref{figure cmd OGLE}). To be able to
compare a CMD constructed with observational data to a CMD constructed
with modelled data, the CMDs have been divided into regions. For both
the 2MASS data and OGLE data, we define a disk region and a bulge
region in the dereddened CMDs as follows:
\begin{eqnarray}
K_{s,0} \le 12.2 & \rm{\ and\ } & (J - K_s)_0 \le 0.25 \rm{\ (for\ 2MASS\
  disk)}\\
I_0 \le 14.1 & \rm{\ and\ } & (V - I)_0 \le 1.4 \rm{\ (for\ OGLE\
  disk)}\\
K_{s,0} \le 10.7 & \rm{\ and\ } & K_{s,0} \ge -10.4 (J - K_s)_0 + 18.9\\
  & & \rm{\ (for\ 2MASS\ bulge)} \nonumber\\
I_0 \le 12.4 & \rm{\ and\ } & I_0 \ge -8.2 (V - I)_0 + 24.0\\
  & & \rm{\ (for\ OGLE\ bulge).} \nonumber
\end{eqnarray}
For the OGLE data we defined additionally a clump region which
contains the red clump stars, as $16.1 \le I_0 \le 12.6$ and $0.8 \le
(V - I)_0 \le 1.8$.\\

The regions are defined on a model with no extinction. Based on the
assumed extinction for a field, these boxes are also reddened, in this
way, these boxes include for every field the same part of the CMD. The
boxes are defined in such a way that there is only a very small
contamination of disk stars in the bulge region and the other way
around. Therefore, if we want to study the bulge parameters, the only
part of interest is the bulge region. The stars in the selected
regions, are used to create histograms. All histograms were created in
one magnitude ($K_s$ for 2MASS and $I$ for OGLE) and not in the
colours. The histograms for the disk and bulge region, which are
probably not affected by incompleteness, are used to
compare the number counts in each bin.\\

\begin{figure}
  \begin{center}
    \resizebox{\hsize}{!}{\includegraphics{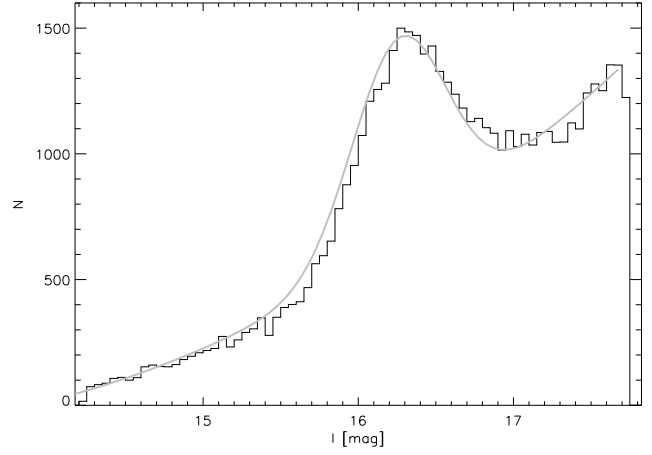}}
    \caption{Example showing how the red clump stars are fitted using
    the method described by \citet{Stanek1998}. The black histogram
    are the stars selected from the ``red clump box'' in the $V - I$ vs
    $I$ CMD and the grey line is the Gaussian fit with underlying
    second degree polynomial.}
    \label{figure fit RC}
  \end{center}
\end{figure}

The histogram for the stars in the red clump region are not used for
number counts, but to derive the red clump magnitude both from
observations and modelled data. To derive the $I_{0,m}$ magnitude for
the red clump, the method described in \citet{Stanek1998} has been
used. This method fits a 2D polynomial to the underlying population,
with on top of this a Gaussian function to determine the magnitude of
the red clump. The formula is given by:

\begin{eqnarray}
n(I_0)  = a & + & b (I_0 - I_{0,m}) + c (I_0 - I_{0,m})^2 \nonumber \\
& + & \frac{N_{RC}}{\sigma_{RC} \sqrt{2\pi}} \exp\left(-\frac{(I_0 -
  I_{0,m})^2}{2\sigma_{RC}^2} \right),
\label{EqRC}
\end{eqnarray}

\noindent with the first tree terms corresponding to the underlying
population with coefficients $a$, $b$, and $c$ for the second degree
polynomial. The last term refers to the red clump population itself,
with parameters corresponding to a Gaussian fit: scaling factor
$N_{rc}$, sigma $\sigma_{RC}$ and peak of the observed red clump stars
$I_{0,m}$. Figure~\ref{figure fit RC} shows a fit using this method to
the $I$-band observations in the bul\_sc40 field. Using this fit to
the red clump, we will compare the peak position of the red clump
stars in the observations to the peak position of the red clump in the
model. The expected magnitude of the red clump $M_I^{\rm RC}$ is often
compared to the peak position in the magnitude distribution of the
red clump in order to estimate the distance to these stars.\\

\subsection{Poisson distributed maximum likelihood}
\label{section poisson maximum likelihood}

For the stars in the selected disk and bulge areas (see
Fig.~\ref{figure cmd 2MASS} and Fig.~\ref{figure cmd OGLE}) the
constructed histograms are used to compare number counts. The typical
errors on the amount of stars in these histogram bins are Poisson
errors ($\sim \frac{1}{\sqrt n_i}$ with $n_i$ the total amount of
stars in bin $i$). Therefore the selected test to perform on these
histogram is a log likelihood test for Poisson statistics
\citep{Eidelman2004}: 
\begin{equation}
\label{equation loglikelihood poisson}
-2\ln(\lambda(\theta)) = 2 \sum_{i=1}^N
\left( \nu_i(\theta) - n_i + n_i  \ln\frac{n_i}{\nu_i(\theta)}
\right).
\end{equation}
In this formula $\theta$ is the set of unknown parameters one wants to
derive, $n = (n_1,\ n_2,\ \ldots, n_N)$ is the data vector containing
the observations with $N$ the number of bins in a histogram. $\nu$ are
the expected values, which are derived from the histograms of the
modelled data and are therefore dependent on $\theta$. When $n_i = 0$,
the last term in Eq.~(\ref{equation loglikelihood poisson}) is
set to zero.\\

For each field, there are 4 histograms (one for
the disk region and one for the bulge region, both for the comparison
with 2MASS and OGLE). When
determining the parameters for the bulge, we only use the histograms
based on the stars in the bulge boxes. This means that for each field,
2 histograms remain. Table~\ref{table selected fields} shows that we
selected 11 fields. Including all this gives us a value $l$ for each
model, which we want to minimise:
\begin{equation}
\label{equation l poisson}
l = \sum_{j=1}^{11} \sum_{k=1}^2 -2\ln(\lambda_{kj}(\theta)).
\end{equation}

\subsection{Gaussian distributed log-likelihood}
\label{section gaussian maximum likelihood}

In addition to a box for the disk and bulge stars, 
Fig.~\ref{figure cmd OGLE} also shows a box in which the red clump stars 
are expected to be. 
Equation~\ref{EqRC} is fitted to the number counts 
in this box, as illustrated in Fig.~\ref{figure fit RC}. The same is done for the model stars.
The parameter of interest is the mean magnitude of the RC, $I_{\rm 0,m}$.

When evaluating calculated models based on the position of the
red clump, we cannot use the Poisson distributed maximum likelihood
(see Sect.~\ref{section poisson maximum likelihood}). In this case
the errors are not Poisson distributed, but normal $N(0,\sigma^2)$. To
determine the appropriateness of a model concerning its position of
the red clump, the log-likelihood function $l$ can be used
\citep{Decin2007}: 
\begin{equation}
\label{equation loglikelihood gauss}
l = \sum_{j=1}^{11}  \ln(\sigma) + \ln(\sqrt{2\pi})
+ \frac{1}{2} 
\left( \frac{I_{\rm obs,j} - I_{\rm model,j}(\theta)}{\sigma}\right)^2.
\end{equation}
To find the
best model, this equation needs to be minimised. As the standard
deviation $\sigma$ we selected 0.025 mag, which is the half of the
bin size used to construct the histograms containing the red clump
stars (see also Sect. \ref{section error bars}). 
\\

\subsection{Minimisation procedure}
\label{section minimisation procedure}
The minimisation procedure used in this work is the
\textit{Broyden-Fletcher-Goldfarb-Shanno (BFGS) method}
\citep{Broyden1970, Fletcher1970, Goldfarb1970, Shanno1970} which can be
used to solve a non-linear optimisation problem. By analysing gradient
vectors, the method constructs an approximated Hessian matrix allowing
a quasi-Newton fitting method to move towards the minimum in parameter
space. The parameters were normalised in the minimisation procedure so
they have an equal influence on the choice of the next set of
parameters.\\

To significantly decrease the amount of computing time, the
simulations made during the minimisation runs, we adopt a shallower
limiting magnitude, at $I = 18$. This does not influence our results
as the stars fainter than this limit have no relevance to our
analysis (see Sect.~\ref{section method}). Because these areas are
chosen well above the completeness limits for both the 2MASS and OGLE
data, we did not include completeness, again to speed up calculations.\\

The model calculations and minimisation procedure have been performed
using Python. The program has been written as such that the 11
different fields, that need to be calculated to evaluate one model
(see Sect.~\ref{section selected fields}), were calculated
simultaneously over 11 different CPU's. We could make use of 18 Dell
PowerEdge PC's with a total of 54 processors of 3400MHz, which makes
it possible to evaluate about 5 models at the same time. In total we
evaluated $\sim 10^5$ models in a time span of about a year. \\

\subsection{Uncertainties on the model parameters}
\label{section uncertainties}

As explained by \citet{Decin2007} the log-likelihood function can be
used to estimate the uncertainties on the model parameters. A model
with a likelihood value $l$ is equally good as the best model
if $\min(l) + N_{\rm crit} > l$. These
critical values $N_{\rm crit}$ follow a $\chi^2_p$-distribution with
$p$ the degrees of freedom \citep{Eidelman2004,Decin2007}. A table
with these values for $p \in [1,20]$ can be found in \citet[][and
  references therein]{Decin2007}.\\


\section{Results}
\label{section results}

\begin{figure}
  \begin{center}
    \resizebox{\hsize}{!}{\includegraphics{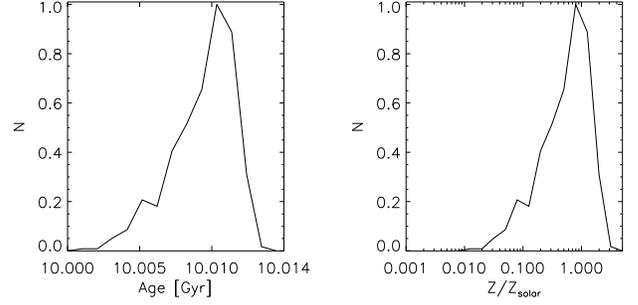}}
    \caption{The left figure shows the SFR normalised to 1 and the
    right figure shows the metallicity distribution normalised to
    1 (based on ZRO2003).}
    \label{figure SFR and AMR}
  \end{center}
\end{figure}

\begin{figure}
  \begin{center}
    \resizebox{\hsize}{!}{\includegraphics{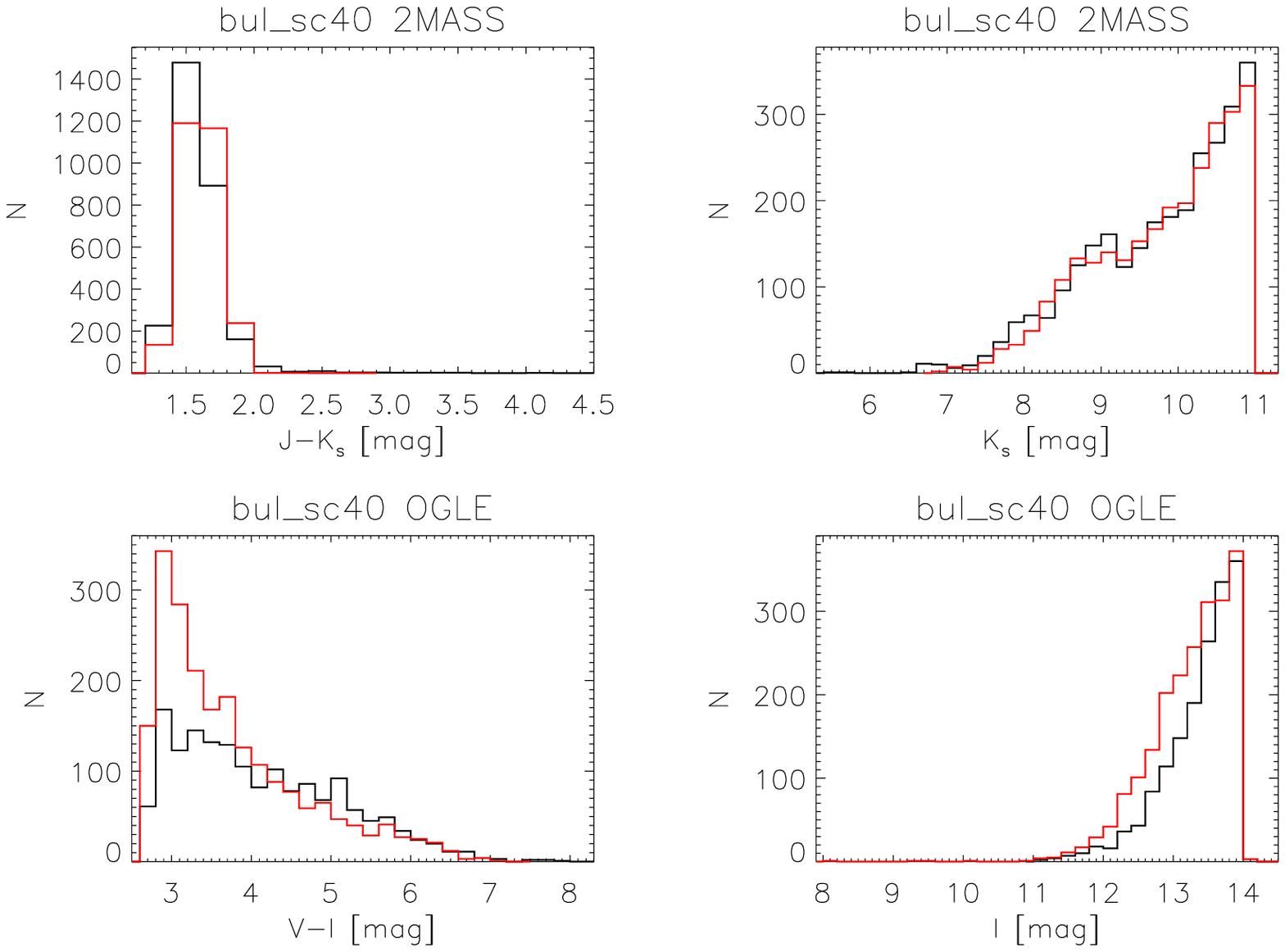}}
    \caption{Histograms showing the distribution of selected bulge
    stars for field bul\_sc40. The black line represents the
    observations, the red line the model. The left panel shows the
    2MASS data, the right panel the OGLE data. The model has the
    following parameters: $R_0 = 8.6$~kpc, $a_m = 2.7$~kpc, $a_0 =
    97$~pc, $\eta = 0.69$, $\zeta = 0.30$, $\phi = 
    20\deg$, and $f_0 = 422.3$. The star formation rate is a 10~Gyr
    burst and the metallicity distribution is based on ZRO2003.}
    \label{figure histo SFR_bulge_zoccali}
  \end{center}
\end{figure}

\begin{figure}
  \begin{center}
    \resizebox{\hsize}{!}{\includegraphics{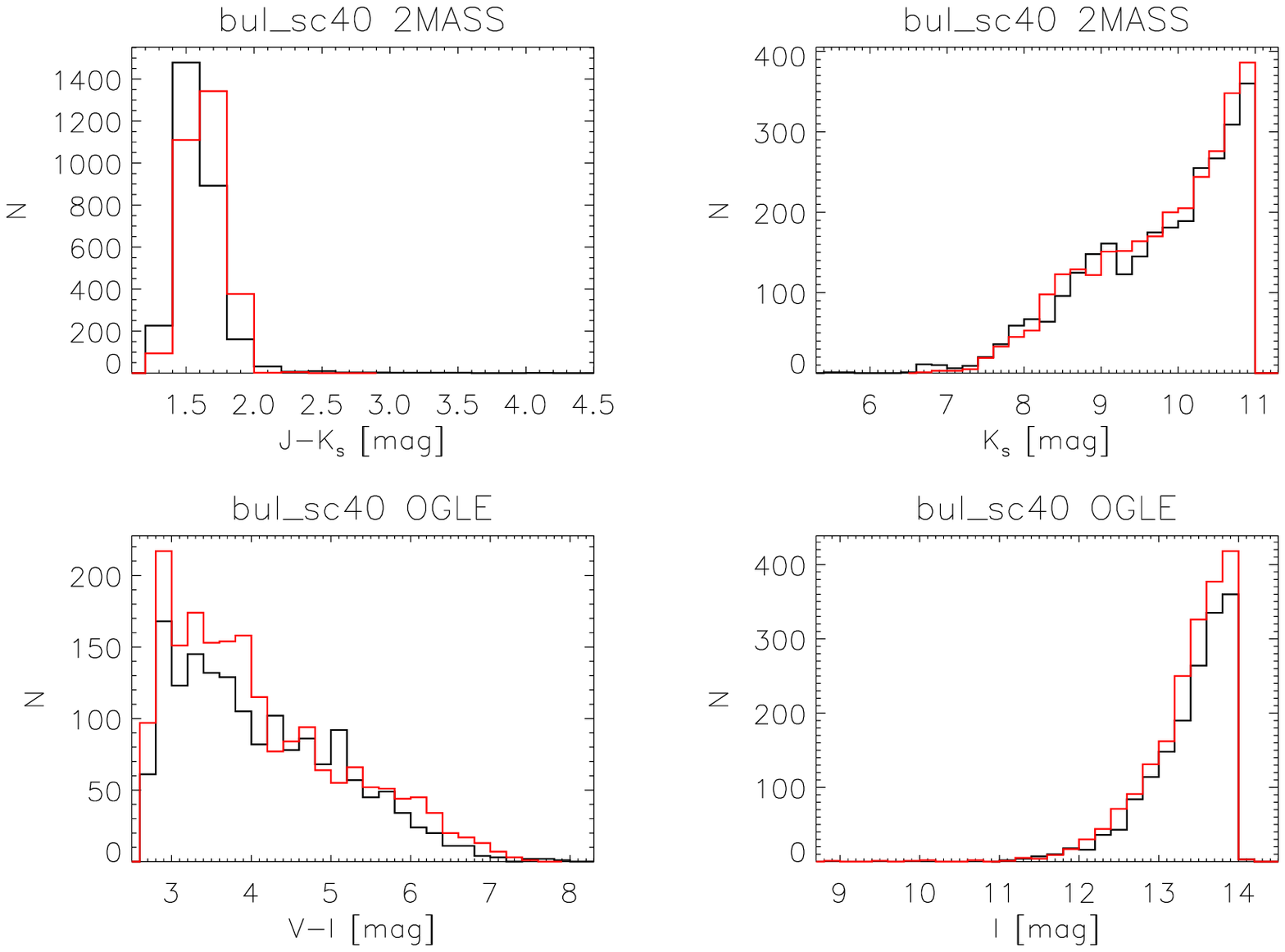}}
    \caption{Histograms showing the distribution of selected bulge
    stars for field bul\_sc40. The black line represents the
    observations, the red line the model. The upper panels shows the
    2MASS data, the lower panels the OGLE data. The model has the
    following parameters: $R_0 = 8.6$~kpc, $a_m = 2.7$~kpc, $a_0 =
    97$~pc, $\eta = 0.68$, $\zeta = 0.30$, $\phi = 
    14\deg$, and $f_0 = 427.3$. The star formation rate is a 10~Gyr
    burst and the metallicity distribution is based on ZRO2003 but
    shifted with 0.3 dex. (This figure is also available in electronic
    form, showing histograms for all the modelled fields.)}
    \label{figure histo SFR_bulge_zoccali_p03 40}
  \end{center}
\end{figure}

\onlfig{9}{
  \begin{figure*}
    \centering
    \resizebox{\hsize}{!}{\includegraphics{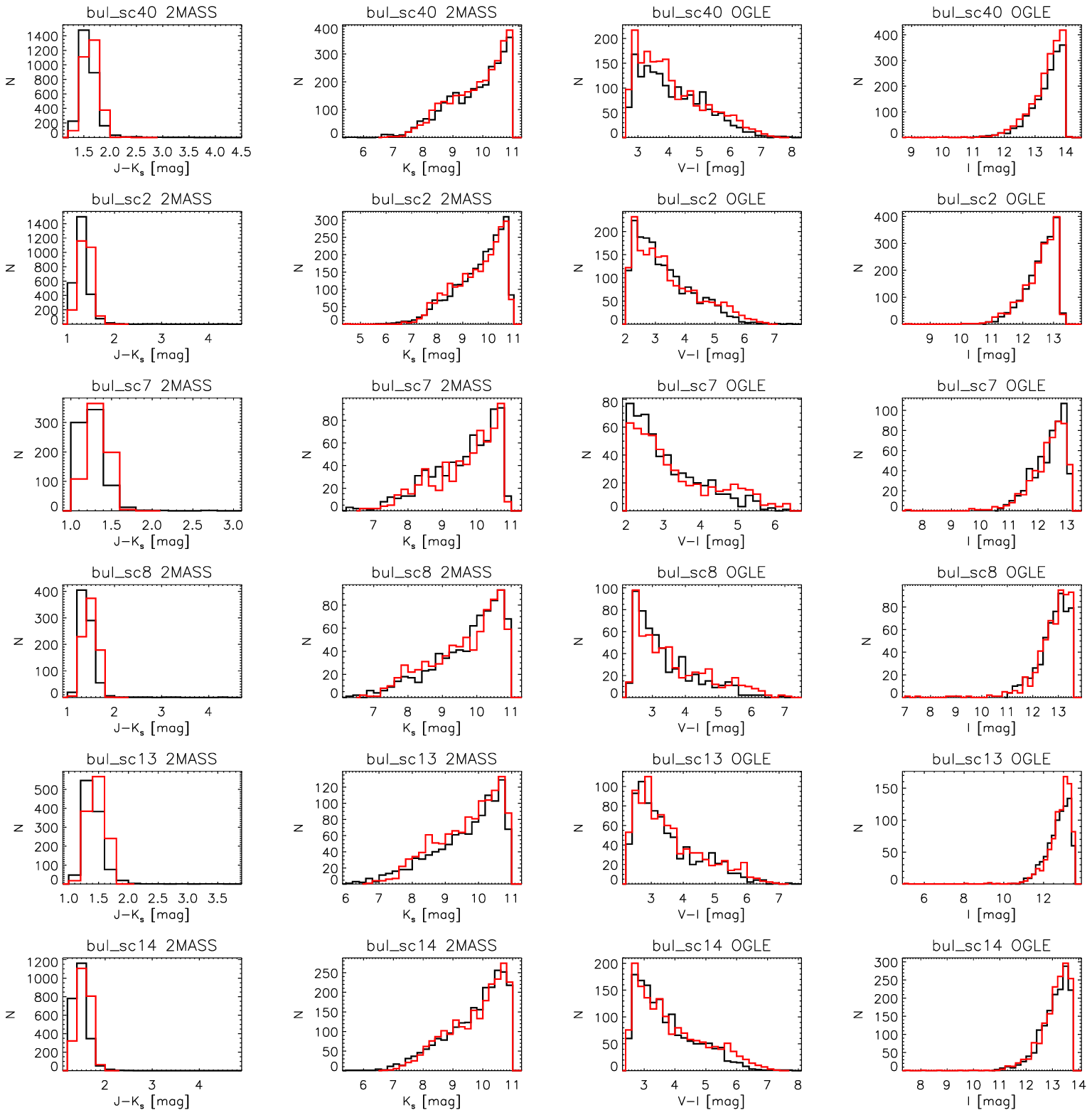}}
    \caption{Histograms showing the distribution of selected bulge
    stars for all the modelled fields. The black line is the histogram
    for the observational data, the red line represents the
    model. For each field there are four panels, the first two panels
    shows the 2MASS data (one for star counts in $J - K_s$ and one for
    the star counts in $K_s$. The other two panels show the star
    counts for the OGLE data ($V - I$ and $I$). The model has the
    following parameters: $R_0 = 8.6$~kpc, $a_m = 2.7$~ kpc, $a_0 =
    97$~pc, $\eta = 0.68$, $\zeta = 0.30$, $\phi = 14\deg$, and
    $f_0 = 427.3$. The star formation rate is a 10~Gyr burst and the
    metallicity distribution is based on ZRO2003 but
    shifted with 0.3 dex.}
    \label{figure histo SFR_bulge_zoccali_p03}
\end{figure*}}

\onlfig{9}{
  \begin{figure*}
    \centering
    \resizebox{\hsize}{!}{\includegraphics{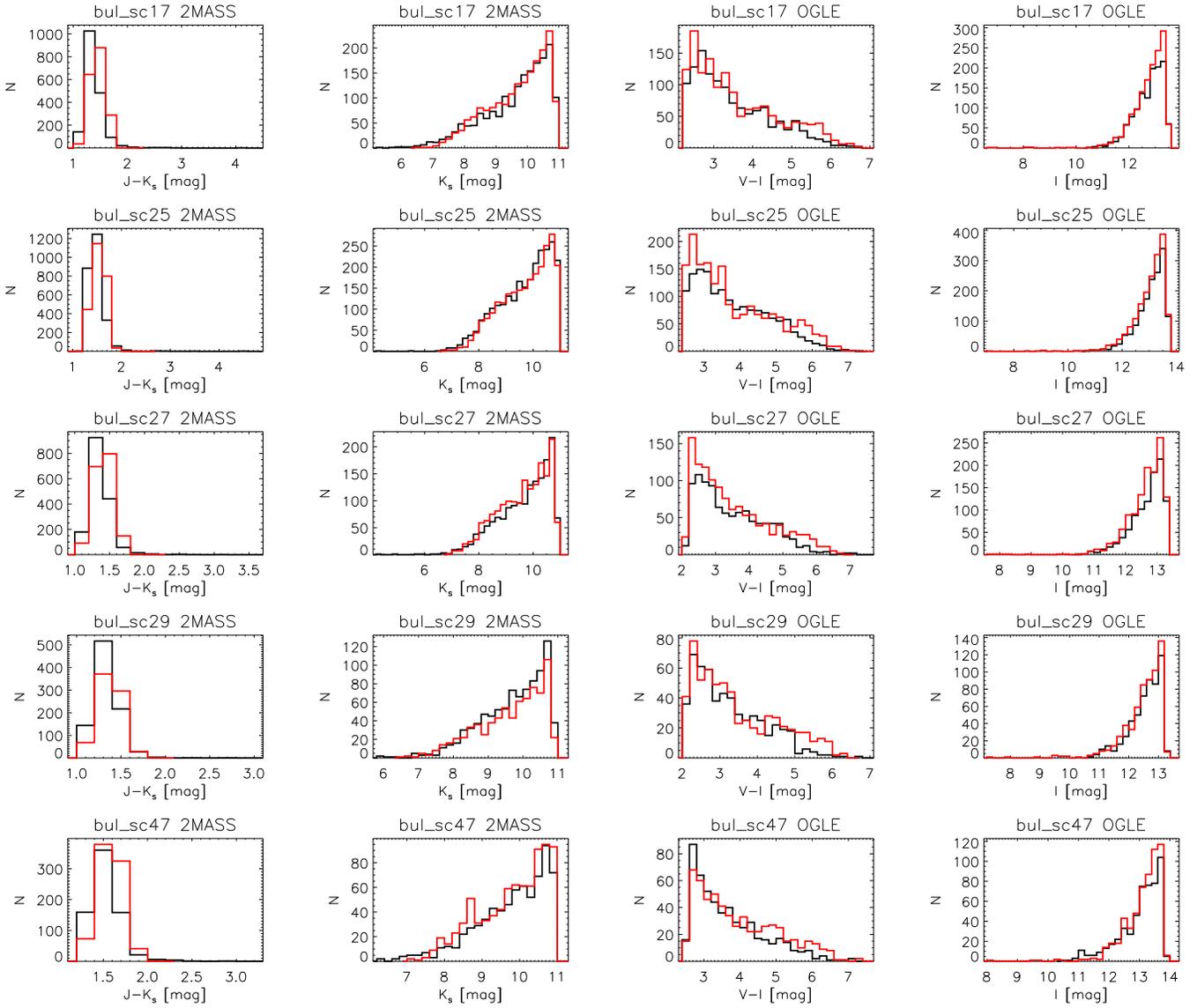}}
    \caption{Continued.}
\end{figure*}}

\begin{table*}
  \caption{Overview of the different models described in the text.}
  \label{table diff SFR}
  \centering
  \begin{tabular}{rrccccccllcr}\hline\hline
    Number & $R_0$ & $a_m$ & $a_0$ & $\eta$ & $\zeta$ & $\phi$ & $f_0$ & SFR & [Fe/H] &
    $l_{\rm bulge}$ & $l_{\rm RC}$\\
    \noindent & [kpc] & [kpc] & [pc] & & & [ \deg ]\\
    \hline
    1& 8.6 & 2.7 & 97 & 0.69 & 0.30 & 20 & 422.3 & ZRO2003, 10~Gyr &
    ZRO2003 & 2\,185 & 10.6\\
    & \\
    2 & 8.6 & 2.7 & 97 & 0.69 & 0.30 & 20 & 422.3 & ZRO2003, 10~Gyr &
    ZRO2003 + 0.1 dex & 1\,768 & 9.2\\
    3 & 8.6 & 2.7 & 97 & 0.69 & 0.30 & 20 & 420.5 & ZRO2003,
    10~Gyr & ZRO2003 + 0.1 dex & 1\,761 & 9.2 \\
    4 & 8.6 & 2.7 & 97 & 0.69 & 0.30 & 20 & 422.3 & ZRO2003, 10~Gyr &
    ZRO2003 + 0.2 dex & 1\,597 & 8.8\\
    5 & 8.6 & 2.6 & 96 & 0.68 & 0.30 & 20 & 422.8 & ZRO2003, 10~Gyr &
    ZRO2003 + 0.2 dex & 1\,576 & 8.4\\
    6 & 8.6 & 2.7 & 97 & 0.69 & 0.30 & 20 & 422.3 & ZRO2003, 10~Gyr &
    ZRO2003 + 0.3 dex & 1\,519 & 9.2\\
    7 & 8.6 & 2.7 & 97 & 0.68 & 0.30 & 14 & 427.3 & ZRO2003,
    10~Gyr & ZRO2003 + 0.3 dex & 1\,442 & 13.3\\
    8 & 8.6 & 2.7 & 97 & 0.69 & 0.30 & 20 & 422.3 & ZRO2003, 10~Gyr &
    ZRO2003 + 0.4 dex & 1\,597 & 8.7\\
    9  & 8.6 & 2.8 & 97 & 0.68 & 0.28 & 20 & 421.6 & ZRO2003,
    10~Gyr & ZRO2003 + 0.4 dex & 1\,560 & 8.3\\
    10 & 8.6 & 2.7 & 97 & 0.69 & 0.30 & 20 & 422.3 & ZRO2003, 10~Gyr &
    ZRO2003 + 0.5 dex & 1\,768 & 8.6\\
    11 & 8.6 & 2.7 & 97 & 0.69 & 0.30 & 21 & 425.5 & ZRO2003,
    10~Gyr & ZRO2003 + 0.5 dex & 1\,717 & 8.5\\
    & \\
    12 & 8.6 & 2.7 & 97 & 0.68 & 0.30 & 14 & 427.3 & ZRO2003,
    7~Gyr & ZRO2003 + 0.3 dex & 3\,517 & 18.0\\
    13 & 8.7 & 2.6 & 96 & 0.68 & 0.28 & 14 & 419.0 & ZRO2003,
    7~Gyr & ZRO2003 + 0.3 dex & 1\,516 & 20.4 \\
    14 & 8.6 & 2.7 & 97 & 0.68 & 0.30 & 14 & 427.3 & ZRO2003,
    8~Gyr & ZRO2003 + 0.3 dex & 2\,247 & 17.0 \\
    15 & 8.7 & 2.6 & 95 & 0.68 & 0.31 & 15 & 406.0 & ZRO2003,
    8~Gyr & ZRO2003 + 0.3 dex & 1\,432 & 9.7\\
    16 & 8.6 & 2.7 & 97 & 0.68 & 0.30 & 14 & 427.3 & ZRO2003,
    9~Gyr & ZRO2003 + 0.3 dex & 1\,735 & 13.0\\
    17 & 8.6 & 2.6 & 96 & 0.69 & 0.29 & 14 & 427.5 & ZRO2003,
    9~Gyr & ZRO2003 + 0.3 dex & 1\,433 & 15.1 \\
    18 & 8.6 & 2.7 & 97 & 0.68 & 0.30 & 14 & 427.3 & ZRO2003,
    11~Gyr & ZRO2003 + 0.3 dex & 1\,657 & 7.1\\
    19 & 8.6 & 2.7 & 97 & 0.68 & 0.30 & 15 & 434.2 & ZRO2003,
    11~Gyr & ZRO2003 + 0.3 dex & 1\,576 & 10.8\\
    20 & 8.6 & 2.7 & 97 & 0.68 & 0.30 & 14 & 427.3 & ZRO2003,
    12~Gyr & ZRO2003 + 0.3 dex & 1\,851 & 25.1\\
    21 & 8.6 & 2.7 & 97 & 0.68 & 0.30 & 13 & 427.5 & ZRO2003,
    12~Gyr & ZRO2003 + 0.3 dex & 1\,564 & 11.9 \\
    22 & 8.6 & 2.7 & 97 & 0.68 & 0.30 & 14 & 427.3 & ZRO2003,
    13~Gyr & ZRO2003 + 0.3 dex & 2\,342 & 7.4\\
    23 & 8.5 & 2.8 & 97 &  0.68 & 0.31 & 16 & 432.3 & ZRO2003,
    13~Gyr & ZRO2003 + 0.3 dex & 1\,696 & 11.5\\
    & \\
    24 & 8.7 & 2.6 & 95 & 0.68 & 0.31 & 15 & 406.0 & vLGO2003 &
    ZRO2003 + 0.3 dex & 2\,190 & 18.7\\
    25 & 10.7 & 3.1 & 86 & 0.73 & 0.30 & 8 & 455.0 & vLGO2003 &
    ZRO2003 + 0.3 dex & 1747 & 120.9\\
    & \\
    26 & 8.7 & 2.6 & 95 & 0.68 & 0.31 & 15 & 406.0 & vLGO2003, &
    ZRO2003 + 0.3 dex & 2\,112 & 18.9\\ 
    & & & & & & & & older than 0.1~Gyr \\
    27 & 9.3 & 2.8 & 92 & 0.66 & 0.30 & 17 & 394.0 & vLGO2003, &
    ZRO2003 + 0.3 dex & 1\,466 & 11.4 \\ 
    & & & & & & & & older than 0.1~Gyr \\
    & \\
    28 & 8.7 & 2.6 & 95 & 0.68 & 0.31 & 15 & 406.0 & vLGO2003, &
    ZRO2003 + 0.3 dex & 3190 & 12.0 \\ 
    & & & & & & & & older than 1.0~Gyr\\
    29 & 9.3 & 2.8 & 92 & 0.66 & 0.30 & 17 & 394.0 & vLGO2003, &
    ZRO2003 + 0.3 dex & 2749 & 17.4 \\
    & & & & & & & & older than 1.0~Gyr\\
    & \\
    30 & 8.7 & 2.6 & 95 & 0.68 & 0.31 & 27 & 406.0 & ZRO2003,
    8 Gyr & ZRO2003 + 0.3 dex & 1426 & 7.3 \\
    \hline
  \end{tabular}
\end{table*}

The SFR and AMR chosen to start the minimisation process with are
based on ZRO2003 and are shown in Fig.~\ref{figure SFR and AMR}. 
The left figure shows a bulge formed by a single star burst
about 10~Gyr ago and the right figure indicates that the majority of
the stars have a metallicity slightly lower than solar ($Z_\odot = 0.019$).\\ 

The start values for the minimisation procedure have been chosen in
accordance with \citet[][see also Table~\ref{table parameters}]{Binney1997}: 
$R_0 = 8.0$~kpc, $a_m = 1.9$~kpc, $a_0 = 100$~pc, $\eta = 0.60$,
$\zeta = 0.40$, $\phi = 20\deg$, and $f_0 = 624.0$. The output of this
minimisation procedure is a model with following parameters: $R_0 =
8.6$~kpc, $a_m = 2.7$~kpc, $a_0 = 97$~pc, $\eta = 0.69$, $\zeta =
0.30$, $\phi = 20\deg$, and $f_0 = 422.3$ 
(see also Table~\ref{table diff SFR}, model 1).\\

The constructed histograms for this model are shown for field
bul\_sc40 in Fig.~\ref{figure histo SFR_bulge_zoccali}. As mentioned
in Sect.~\ref{section method} we only used the histograms constructed
with the $I$ and $K_s$ magnitude to compare the model with the
observations. The histograms in $V - I$ and $J - K_s$ illustrate the fit
between the model and the data for the colours. The minimisation
procedure uses only the stars in the bulge box. The stars in the disk
box were used prior to the minimisation process to set the scale
parameter for the disk stars and to check the contamination of disk
and halo objects in the ``bulge box''. For field bul\_sc40 adopting the
same model as in Fig.~\ref{figure histo SFR_bulge_zoccali}, we find
that out of a total of 2\,731 objects in the bulge box for the 2MASS
data only 127 originate from the disk and halo. Concerning the 2\,110
objects in the bulge box for the OGLE data, only 105 originate from the
disk and halo. Since this is already a very modest fraction (about 5\%
for both CMDs), possible errors in our representation of the disk and
halo geometry are expected to have a negligible impact on the star
counts inside the bulge box.\\

Fig.~\ref{figure histo SFR_bulge_zoccali} shows the distribution of
stars for the observations and the model for the selected bulge stars
for field bul\_sc40. Also for all the other fields this model
overestimates the amount of bulge stars in the 2MASS data and
underestimates the amount of bulge stars in the OGLE data, therefore
this model can still be improved, but not by adapting the parameters
already in the minimisation procedure.\\

\subsection{Varying the metallicity distribution}
\label{section varying the metallicity distribution}

One of the parameters not in the minimisation procedure is the
metallicity. The metallicity distribution in ZRO2003, which has been
used for the previous model calculations, is based on photometric
results and is slightly less metal-rich in comparison with metallicity
distributions derived from spectroscopic results from e.g. \citet{McWilliam1994}, 
\citet{Sadler1996}, and \citet{Ramirez2000} (see Fig.~14 in
ZRO2003). Therefore, we shifted the metallicity distribution
towards a more metal-rich population. Already when we shift [Fe/H]
with 0.1 dex our fit improves (see model 2 in Table~\ref{table
    diff SFR}). The metallicity distribution has been shifted with
  +0.1, +0.2, +0.3, +0.4, and +0.5 dex (see respectively models 2, 4,
  6, 8, and 10 in Table~\ref{table diff SFR}). We obtained the best
fit with a shift of +0.3 dex for [Fe/H]. New minimisation
  procedures have been set up using these shifted metallicity
  distributions in order to test the effect of the change in
  metallicity distribution on the derived parameters. The minimisation
  procedure was started from the result of the previous minimisation
  process (see model 1 in Table~\ref{table diff SFR}). The results are
  listed in Table~\ref{table diff SFR} (see models 3, 5, 7, 9, and
  11). Reminimising the models revealed no differences concerning the
  metallicity distribution shift, again the best results are found
  using a shifted metallicity distribution of +0.3 dex with respect to
  ZRO2003. Concerning the parameters describing the geometry of the
  bulge and its bar, only $\phi$ and $f_0$ change. These changes are
  small and fall within the expected error bars (see
  Sect.~\ref{section error bars}).  Fig.~14 in ZRO2003 shows that a
shift with 0.3 dex in [Fe/H] is consistent with the noticed
differences between the metallicity distribution they derived and
metallicity distributions derived based on spectroscopy. In fact,
  \citet{Zoccali2008} have recently rederived the bulge metallicity
  distribution, based on high resolution spectroscopy for a large
  sample. They find metallicities systematically higher than in ZRO03
  by about 0.2 dex, which is perfectly in line with what we obtain in
  this paper.\\

\subsection{Varying the age distribution}
\label{section varying the age distribution}

The star formation rate is also a input to the model (see
Sect.~\ref{section SFR and AMR}). Therefore we also tested other
SFR's. First we start with shifting the star burst to a different age
in steps of 1~Gyr from a star burst 7~Gyr ago to a
  star burst 13~Gyr ago. Table~\ref{table diff SFR} gives an
  overview of these models when no new minimisation procedure is
  started (see models 12, 14, 16, 18, 20, and 22). Amongst these
  models, there is no better model than model 7. When a minimisation
  procedure is set up, using the different age distributions and the
  parameters describing the geometry as derived in model 7 as start
  values, we find models 13, 15, 17, 19, 21, and 23 (see
  Table~\ref{table diff SFR}). Three of these models are equally good
  based on $l_{\rm bulge}$: model 15 with a star burst of 8~Gyr is the
  so-called best model, but models 17 (star burst of 9~Gyr ago) and 7
  (star burst 10~Gyr ago) are equally good and one cannot distinguish
  between these three models. Concerning the parameters describing the
  geometry, the differences seen between these three models, fall
  within the expected error bars (see Sect.~\ref{section error bars}).\\

Changing the age of the star burst is not the only possibility to
alter the SFR. Sect.~\ref{section SFR and AMR} describes that most
likely there are also intermediate age stars in the GB and maybe also
some younger stars. In order to check these possibilities, we used the
SFR given in vLGO2003 (see their Fig.~28). The metallicity
distribution used is the metallicity distribution of ZRO2003 shifted
with 0.3 dex. Young stars as well as intermediate age stars are
included in the SFR by vLGO2003. To check the possibility of an
intermediate age population on top of the dominant old population, the
age distribution of vLGO2003 has been used, but without the
stars younger than 0.1~Gyr and a SFR without the stars younger than
1~Gyr. The metallicity distribution used in these scenarios is again
the shifted distribution found by ZRO2003. The results are again shown
in Table~\ref{table diff SFR}.\\

Model 24 in Table~\ref{table diff SFR} shows that we do not get a
significantly better fit if we use the SFR by vLGO2003 and the shifted
metallicity distribution by ZRO2003. Because the SFR has changed
drastic, and therefore also the colour distribution of the stars in
the resulting models, a new minimisation procedure is necessary to
check if no better model can be found using the SFR by vLGO2003. The
previous best model found (model 15), has been chosen to start
the minimisation procedure with. This resulted in the following model:
$R_0 = 10.7$~kpc, $a_m = 3.1$~kpc, $a_0 = 86$~pc, $\eta = 0.73$,
  $\zeta = 0.30$, $\phi = 8\deg$, and $f_0 = 455.0$ (model 25 in
  Table~\ref{table diff SFR}). All parameters except $R_0$ and $f_0$
fall within the expected error bars of the best model with a
star burst of 10~Gyr as SFR (see Sect.~\ref{section error
    bars}). The differences seen between $R_0$ and $f_0$ are huge and
do not fall within the previously defined error bars. $l_{\rm
    bulge}$ Additionally, shows that this model is not significantly
  better than model 15.\\

Also for the two SFR's including only intermediate age stars based on
the SFR by vLGO2003 a new minimisation procedure was set up. Using the
SFR by vLGO2003 for stars older than 0.1~Gyr, the minimisation
procedure ended with the following model: $R_0 = 9.3$~kpc, $a_m =
2.8$~kpc, $a_0 = 92$~pc, $\eta = 0.66$, $\zeta = 0.30$, $\phi =
17\deg$, and $f_0 = 394.0$. All parameters found using this
SFR are within the error bars derived on model 15 (see
  Sect.~\ref{section error bars}). Using the SFR for stars older than
1~Gyr, the  minimisation procedure ended with the same model: $R_0
  = 9.3$~kpc, $a_m = 2.8$~kpc, $a_0 = 92$~pc, $\eta = 0.66$, $\zeta =
  0.30$, $\phi = 17\deg$, and $f_0 = 394.0$, which then of course
falls also within the error bars of model 15. Based on $l_{\rm bulge}$,
these two later models are not as good as model 15.\\

\subsection{Error bars}
\label{section error bars}

To derive the error bars on these model parameters, we used 
the method described in Sect.~\ref{section uncertainties}. 
For each
parameter we explored the parameter space around the value found by
the minimisation procedure and rounded the found parameters within the
found error bars. 
All models that fall within
the critical value of $N_{\rm crit} = 14.07$ from our model are
equally good. This results in the following error bars for model
  7: $R_0=8.6\pm^{0.15}_{0.10}$~kpc, $a_m=2.7\pm^{0.04}_{0.08}$~kpc,
$a_0=97\pm^{1.1}_{3.5}$~pc, $\eta=0.68\pm^{0.01}_{0.07}$,
$\zeta=0.30\pm^{0.01}_{0.01}$, $\phi=14\deg\pm^{5.6}_{9.6}$, and
$f_0 = 427.3\pm^{10.8}_{25.7}$.\\


These error bars have been used in the previous section to
  determine whether parameters between different models fall within
  the same range. As model 15 was found to be the best model, we also
  calculated the error bars for this model.
We obtain: 
  $R_0=8.7\pm^{0.57}_{0.42}$~kpc, $a_m=2.5\pm^{1.73}_{0.13}$~kpc,
  $a_0=95\pm^{7.5}_{12.8}$~pc, $\eta=0.68\pm^{0.04}_{0.19}$,
  $\zeta=0.31\pm^{0.06}_{0.04}$, $\phi=15\pm^{13.3}_{12.7}$, and
  $f_0=406.0\pm^{40.4}_{167.3}$. 
The error bars on this
  model are somewhat larger than the error bars found for model 7.\\

Based on the outcome of the minimisation procedures, the error
  bars on the determination of the magnitude of the red clump stars
  in the $I$-band have been determined. This resulted in an average
  error of 0.03 mag which is consistent with the chosen value in
  Eq.~(\ref{equation loglikelihood gauss}).


\section{Red Clump stars}
\label{section RC stars}

The red clump stars can be used as a distance indicator as well as an
estimator to the angle $\phi$ between the bar and the Sun-centre
line (see also Sect.~\ref{section parameters}). Using the red clump
stars, a new minimisation procedure as described in Sect.~\ref{section
  minimisation procedure} was set up. In this scenario, there are only
two variable parameters: $R_0$ and $\phi$, the other parameters have
been kept constant ($a_m = 2.6$~kpc, $a_0 = 95$~pc, $\eta = 0.68$,
$\zeta = 0.31$, and $f_0 = 406.0$). An age of 8~Gyr and a
metallicity distribution based on ZRO2003 and shifted with 0.3 dex has
been used (which is consistent with the best model found, 15). 
The minimisation process started with $R_0 = 8.7$~kpc and $\phi
= 14\deg$ and ended with $R_0 = 8.7$~pc and $\phi = 26.7\deg$. 
This model is also listed in Table~\ref{table diff  SFR}.\\

Using the 
method to derive error bars on these parameters, with
$N_{\rm crit} = 5.99$, the following error bars were calculated: 
$R_0 = 8.7\pm^{0.62}_{0.54}$~kpc and $\phi = 26.7\deg\pm^{4.5}_{26.7}$. 
Therefore the model derived based on the red clump stars for $R_0$ and
$\phi$ falls within the previously defined error bars for the model
derived on the stars located in the bulge box. Although this
  model has a smaller $l_{\rm bulge}$ and $l_{\rm RC}$, it is not
  significantly better than the previous found best model (see model
  15 in Table~\ref{table diff SFR}.) \\

In Sect.~\ref{section results} other SFR's have been tested (see
Table~\ref{table diff SFR}). Also for these other models, we
compared the red clump stars. Based on the red clump stars,
Table~\ref{table diff SFR} shows that the models with a star burst of
10~Gyr as SFR but with shifted metallicity distributions are equally
good, except for model 7 which has a significantly worse $l_{\rm RC}$. 
There is based on the critical value $N_{\rm crit}$ no difference
between these models. Concerning the models with a SFR based on
vLGO2003 the comparison with the red clump stars shows that our fit
could not be improved adapting this SFR.\\

\begin{figure}[t]
  \begin{center}
    \resizebox{\hsize}{!}{\includegraphics{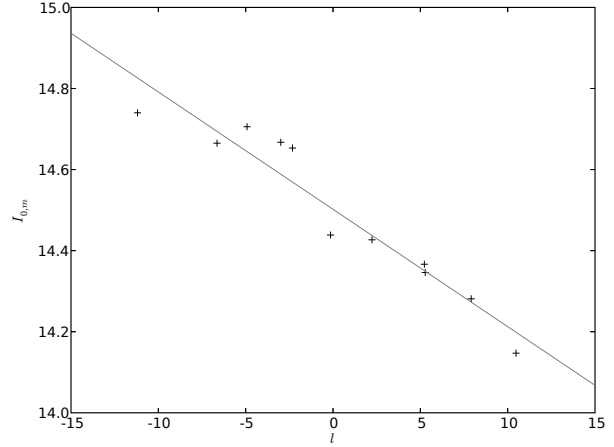}}
    \caption{Peak positions $I_0$  of the red clump stars corrected
    for extinction in function of the longitude $\ell$.}
    \label{figure calibrate MI}
  \end{center}
\end{figure}

The results of this minimisation, and in particular the $R_0$
value, are clearly associated to the set of stellar models that we
used. Models with a different absolute magnitude of the red clump, for
instance, would lead to a different GC distance. \citet{Girardi1998}
and \citet{Girardi2001} claim the set of stellar models we are
using \citep{Girardi2000} do reproduce the very accurate $M_I^{\rm RC}$
value derived from Hipparcos; however their claims are tight to the
specific choice of SFR \citep[from][]{Rocha-Pinto2000b} and AMR
\citep[from][]{Rocha-Pinto2000a} employed in modelling the Solar
Neighbourhood. Future revisions of these functions may lead to
different results.\\

Several other works have already tried to access the bulge orientation
and GC distance using red clump stars, employing different data,
extinction corrections, and assumptions about the intrinsic magnitude
difference between the local (Hipparcos) and the bulge red clump
(i.e. the population correction, $\Delta
M_\lambda^{\rm RC}=M_\lambda^{\rm RC}({\rm Hipp}) - M_\lambda^{\rm
  RC}({\rm galaxy})$).
>From the entries in Table~\ref{table parameters}, one notices that our
best $R_0$ value agrees with the distance determinations to the bulge
from \citet{Paczynski1998} and \citet{Stanek1998}, which were also
derived from $I$-band data from OGLE. However, these are significantly
larger than those derived from \citet{Babusiaux2005} and
\citet{Nishiyama2006} using near-infrared data.\\

In order to allow any future comparison between our $R_0$ determination
and those from other authors, we provide below a representative value
of the absolute red clump magnitude of our models.\\

Since TRILEGAL simulates the entire population along a given
line-of-sight, our simulated LFs for the red clump box include effects
like the dispersion in the distances of bulge stars, geometrical
factors, and the contamination by disk and halo stars. As a
consequence, our effective $M_I^{\rm RC}$ values do vary slightly from
region to region even though the bulge stellar population which is
being simulated is everywhere the same. Moreover, the 
Galactic centre line-of-sight has not been simulated, and a representative
value of $M_I^{\rm RC}$ has to be inferred from the values distributed
across the bulge. Figure~\ref{figure calibrate MI} shows the observed
peak positions of the red clump, $I_{0,m}$ as a function of longitude
$\ell$. These peak positions are corrected for extinction using the
extinction values given in Table~\ref{table selected fields}. The full
line is a linear least squares fit through these peak positions and is
given by $I_{0,m} = (14.502 \pm 0.018) + (-0.029 \pm 0.03)\ell$. 
Therefore the peak position $I_{0,m}$ in the GC ($\ell=0$)
is given by $14.502\pm0.018$. This is related to the absolute red
clump magnitude by
\begin{equation} 
\label{equation MI}
M_I^{\rm RC} = I_{0,m} - (5 \log R_0 - 5).
\end{equation}
Using the distance to the GC $R_0=8.7\pm^{0.57}_{0.42}$~kpc derived from
our minimisation procedures for an age of 8~Gyr, and the peak
position $I_{0,m}$ inferred for the GC, this equation gives
$M_I^{\rm RC} = -0.196\pm^{0.143}_{0.106}$. This value is consistent with the
$M_I^{\rm RC} = -0.185\pm0.016$ used by \citet{Paczynski1998}, and with the
$M_I^{\rm RC} = -0.23\pm0.03$ used by \citet{Stanek1998}, in their GC
distance determinations\footnote{We recall that their $M_I^{\rm RC}$ values
were derived from the Hipparcos sample of red clump stars with good
parallax measurements, without applying any population correction
between the Hipparcos and bulge samples. \citet[their Table 4]{Girardi2001}
 estimated that this correction is smaller than 0.1 mag.}. 
The value is also consistent with the recent determination of 
$M_I^{\rm RC} = -0.22\pm0.03$ by \citet{Groenewegen2008} from 
an analysis based on the revised Hipparcos parallaxes.
This explains why their distance determinations are consistent
with ours despite the quite different methods employed.\\

On the other hand, our preferred $R_0$ value disagrees with the
\citet{Babusiaux2005} and \citet{Nishiyama2006} distance
determinations based on near-infrared data for the red clump.
These works find $R_0$ values close to 7.5~kpc, which means a
disagreement at the level of 0.3 mag in distance modulus. The origin
of this discrepancy is not understood at the moment.

It is worth remarking that our favoured $R_0$ value of 8.7~kpc
comes from both the red clump and the bulge box, the latter consisting
mainly of upper-RGB and TP-AGB stars. These stellar groups are quite
independent in their photometric properties.  Our $R_0$ value is also
consistent with the determinations based on the RR~Lyrae and Mira
variables.  Although it disagrees with with the BH geometrical
distance derived from \citet{Eisenhauer2003,Eisenhauer2005} it does
agree with the recent determination by \citet{Ghez2008}.  \\


\rm

\section{Summary and conclusions}
\label{section conclusions}

In this paper we have studied the characteristics of the Galactic
bulge based on CMD comparisons between OGLE and 2MASS data an the
results from the galactic model TRILEGAL \citep{Girardi2005}. Several
star formation rates and metallicity distributions have been
tested. Over all these different SFR's and metallicity distributions
the model parameters give within the error bars the same results
except for the angle $\phi$ between the Sun-centre line and the
Bar. Concerning the other parameters we found a distance to the
Galactic centre $R_0 = 8.7\pm^{0.57}_{0.43}$~kpc, a scale length for
the Bulge $a_m$ of $2.5\pm^{1.73}_{0.16}$~kpc, an inner truncation
length for the Bulge $a_0$ of $95\pm^{7.5}_{12.8}$~pc, and a scaling
factor $f_0 = 406.0\pm^{40.4}_{167.3}$. Concerning the
characteristics of the Bar, we found the ratio of the major axis to be
$1 : 0.68\pm_{0.05}^{0.19} : 0.31\pm_{0.06}^{0.04}$ and an angle
$\phi$ of $15\deg\pm_{13.3}^{12.7}$. The largest scatter in the found
model parameters is found for the angle $\phi$. Nevertheless the found
values are all rather small and around $15 \sim 20\deg$. The present best model
of the bulge will be provided as the default in the interactive web
interface to TRILEGAL ({\tt http://stev.oapd.inaf.it/trilegal}).\\

If we compare our results to the results found in the literature and
listed in Table~\ref{table parameters} our parameters fall within the
listed range of parameters except for the axial ratio $\eta$. 
Concerning the distance to the GC, our value resembles the
values found by \citet{Paczynski1998} based on red clump stars and
\citet{Groenewegen2005} based on Mira variables. Our results disagree
with e.g. \citet{Udalski1998} based on red clump stars in the GB. If
we compute the mean observed magnitude $I_0$ from the distance modulus
listed in \citet{Udalski1998} using $M_I^{\rm RC}$ from \citet{Stanek1998} 
as they did, than this is consistent with the observed mean magnitude
we derived in this work. The other two studies deriving the distance
to the GC based on red clump stars which are inconsistent with this
work are \citet{Babusiaux2005} and \citet{Nishiyama2006}. These two
studies are consistent with each other and use a very similar method:
both define an ``extinction-free'' magnitude. It is not exactly clear
why the results of these methods are not consistent with our
results. As these studies are performed using $K$-band observations,
we can not compare the expected magnitude as we could do for
\citet{Udalski1998}. For the scale length of the bulge, we found the
same value as \citet{Freudenreich1998} and similar values as
\citet{Sevenster1999a} and \citet{Bissantz2002}. The inner truncation
length for the Bulge is similar to the two values listed in
Table~\ref{table parameters} by \citet{Binney1997} and
\citet{Bissantz2002}. The value $\eta$ we found is higher than the
values found in the literature. Concerning the other ratio of the major
axis $\zeta$ we found similar results as \citet{Whitelock1992},
\citet{Dwek1995}, and \citet{Stanek1997}. For the last parameter
$\phi$, we found values consistent with the group of studies finding
lower values for $\phi$ \citep[see e.g.][]{Dwek1995, Binney1997,
  Stanek1997, Bissantz2002, Robin2003, Merrifield2004, Babusiaux2005,
  Lopez-Corredoira2005, Rattenbury2007}. As already mentioned by
\citet{Groenewegen2005} the different results found in the literature with
respect to the angle between the bar and the Sun-centre line
\citep[see e.g. our results in contrast to][]{Whitelock1992,
  Sevenster1999a, Groenewegen2005, Lopez-Corredoira2007} could
originate from a different spatial distribution of the stellar
population these studies trace.\\

We also tested different star formation rates and metallicity
distributions. A model with a star burst of 8~Gyr based on ZRO2003
gave the best results together with a metallicity distribution also
based on ZRO2003 but shifted by 0.3 dex, although we could find no clear
discrepancy between this model and the models with a star burst 10~
Gyr ago could be found. We could not improve our fits
including intermediate and/or young stars if we compared both the
stars we selected in the defined bulge box and the red clump
stars. Therefore we conclude that based on 2MASS and OGLE data
comparisons with the TRILEGAL model we find best results for a
population of 8~Gyr. Nevertheless it is still possible that there are
also intermediate age stars located in the GB, although we could not
trace them using our described method \citep[see e.g.][]{Groenewegen2005,Uttenthaler2007}.\\

Using the peak positions of the red clump stars and the distance to
the Galactic centre retrieved by our minimisation procedures, we found
that $M_I^{\rm RC} = -0.170 \pm_{0.066}^{0.069}$. This value is
consistent with recent values found in the literature, although it
resembles best the values found by \citet{Girardi2001} using red clump
stars in the solar neighbourhood ($M_I^{\rm RC} = -0.171$) or using
red clump stars in Baade's Window if $\alpha$-enhancement is taken
into account ($M_I^{\rm RC} = -0.161$).\\

\begin{acknowledgements}
E.V. would like to thank the system administration group of the
Institute for Astronomy in Leuven and in special E. Broeders for the
helpful discussions with respect to the implementation of the
described minimisation algorithms. L.G. acknowledges partial
  support by the University of Padova (Progetto di Ricerca di Ateneo
  CPDA052212). This publication makes use of data products from the
Two Micron All Sky Survey, which is a joint project of the University
of Massachusetts and the Infrared Processing and Analysis
Center/California Institute of Technology, funded by the National
Aeronautics and Space Administration and the National Science Foundation.
\end{acknowledgements}



\end{document}